\documentclass[pre,twocolumn,superscriptaddress,showpacs,preprintnumbers,amsmath,amssymb]{revtex4}
\usepackage{graphicx,epsfig}
\usepackage{dcolumn}
\usepackage{bm}
\usepackage{psfrag}
\usepackage{plain}
\usepackage{tabularx}
\usepackage[latin1]{inputenc}
\usepackage{amsmath}
\begin{document}

\title{Transport on exploding percolation clusters}

\author{Jos\'e S. Andrade Jr.}
\affiliation{Departamento de F\'{\i}sica, Universidade Federal
do Cear\'a, 60451-970 Fortaleza, Cear\'a, Brazil}

\author{Hans J. Herrmann}
\affiliation{Departamento de F\'{\i}sica, Universidade Federal
do Cear\'a, 60451-970 Fortaleza, Cear\'a, Brazil}
\affiliation{Computational Physics, IfB, ETH-H\"onggerberg,
Schafmattstrasse 6, 8093 Z\"urich, Switzerland}

\author{Andr\'e A. Moreira}
\affiliation{Departamento de F\'{\i}sica, Universidade Federal
do Cear\'a, 60451-970 Fortaleza, Cear\'a, Brazil}

\author{Cl\'audio L. N. Oliveira}
\affiliation{Departamento de F\'{\i}sica, Universidade Federal
do Cear\'a, 60451-970 Fortaleza, Cear\'a, Brazil}

\begin{abstract}
  We propose a simple generalization of the explosive percolation
  process [Achlioptas {\it et al.}, Science 323, 1453 (2009)], and
  investigate its structural and transport properties. In this model,
  at each step, a set of $q$ unoccupied bonds is randomly chosen. Each
  of these bonds is then associated with a weight given by the product
  of the cluster sizes that they would potentially connect, and only
  that bond among the $q$-set which has the smallest weight becomes
  occupied. Our results indicate that, at criticality, all finite-size
  scaling exponents for the spanning cluster, the conducting backbone,
  the cutting bonds, and the global conductance of the system, change
  continuously and significantly with $q$. Surprisingly, we also
  observe that systems with intermediate values of $q$ display the
  worst conductive performance. This is explained by the strong
  inhibition of loops in the spanning cluster, resulting in a
  substantially smaller associated conducting backbone.
\end{abstract}


\maketitle

The quest for a percolation paradigm that displays a first-order
transition has been the focus of intensive research in Statistical
Physics~\cite{Chalupa79}. Very recently, an extension of the
traditional percolation model has been proposed that displays a sudden
transition of the order parameter as a function of the bond occupation
probability~\cite{Achlioptas09}. In this so-called ``explosive
percolation'' (EP) process, bonds are selected to be occupied in
accordance with a product rule that favors the growth of smaller
clusters over large ones. Interestingly, although this model presents
an abrupt transition when applied to different network
topologies~\cite{Ziff09,Radicchi09,Friedman09}, its critical phase
reveals signatures of a typical continuous transition, as for example,
a power-law distribution of cluster sizes~\cite{Radicchi09}.  This
controversial nature of the EP transition has been recently addressed
in Ref.~\cite{Costa10}, where it has been argued through analytical
arguments that the transition may actually be continuous for the case
of random graphs. As a potential application, the EP model has been
linked to the growth dynamics of Protein Homology
Networks~\cite{Rozenfeld10}. Abrupt transitions have also been
recently observed for other percolation-like
models~\cite{DSouza10,Cho10}.  For instance, it has been shown that,
at least in the limit of infinite dimensionality, the absence of loops
represents an important ingredient to obtain an explosive percolation
process~\cite{Moreira10}.  Subsequently, a study on regular lattices
in which the growth of the largest cluster is systematically
controlled during the percolation process, provided unambiguous
evidence for a first-order phase transition~\cite{Araujo10}.

\begin{figure*}[t]
\begin{center}
\includegraphics*[width=18.0cm]{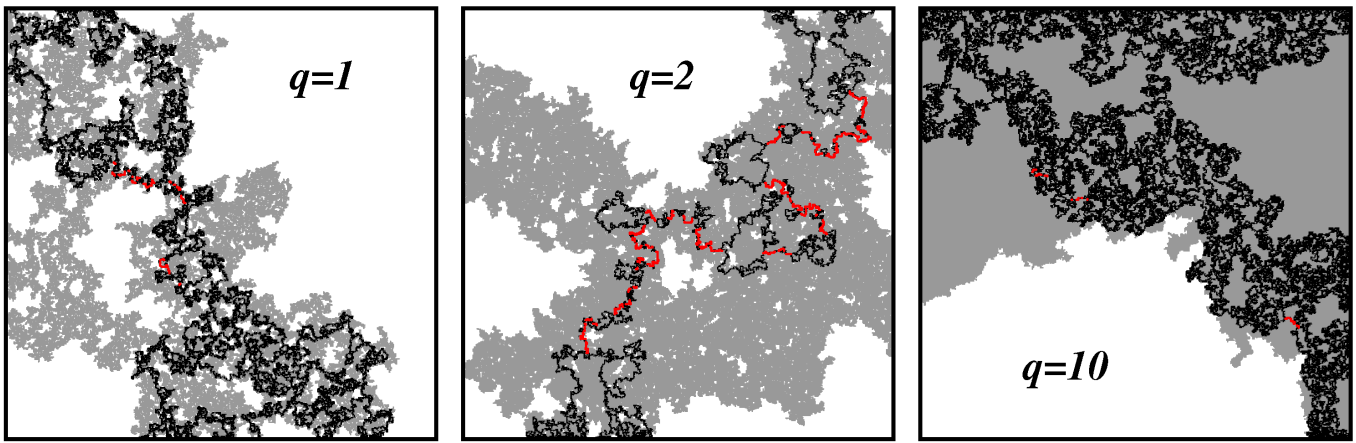}
\end{center}
\caption{(color online) Typical clusters obtained from the EP
  percolation process with $q$-links on a square lattice of size
  $L=512$. In this model, at each step a set of $q$ unoccupied bonds
  is randomly chosen. Each of these bonds has a weight given by the
  product of the cluster sizes that they would potentially connect.
  Only the bond with minimum weight in the set becomes occupied. The
  process continues in sequential steps, and is halted when a cluster
  forms connecting the top and the bottom of the lattice.  Here we
  show clusters for $q=1$ (traditional bond percolation model), $q=2$
  (Achlioptas model~\cite{Achlioptas09}), and $q=10$. The spanning
  cluster appears in gray, the conducting backbone in black, and the
  cutting bonds in red. For $q=10$ the spanning cluster occupies
  almost the entire lattice and the conducting backbone constitutes a
  large fraction of the whole system. Interestingly, the number of
  cutting bonds is significantly larger in the case $q=2$.}
\label{f.pic}
\end{figure*}

The problem of transport in disordered media has been extensively
studied under the traditional paradigm of percolation
theory~\cite{Stauffer92,Sahimi94}. Metal-insulator
transitions~\cite{Last71} and flow through porous
media~\cite{Stanley84} are just two among several physical phenomena
that have been successfully modeled under the framework of percolation
as a characteristic structural second-order phase transition. At the
critical point, percolation systems typically display a conductance
$\sigma$ that decays with the lattice size as $\sigma\sim{L}^{-\mu}$,
with an exponent $\mu$ that depends exclusively on the lattice
dimensionality~\cite{Stauffer92,Sahimi94}. Here we investigate the
structural and transport properties of EP networks built on a square
lattice. In our simulations a generalization of the product rule
proposed in Ref.~\cite{Achlioptas09} is adopted, where at each step a
bond is selected from a random set of $q$ unoccupied bonds.
Strikingly, our results suggest that the scaling exponents associated
with structural and transport quantities change continuously with the
parameter $q$. We observe that the abrupt behavior of the percolation
transition is stressed as $q$ increases, reaching a limiting regime
for large $q$ values, similar to the model where the selected
connection is the one with the smallest possible weight in the entire
network~\cite{Manna09}. Moreover, the global conductivity of the
system exhibits a very slow decrease with system size for large values
of $q$, with an exponent that approaches zero, while at intermediate
values the exponent is maximum, i.e., it decreases faster.

The present model is implemented on a square lattice of size $L$.
Initially all the bonds of the lattice are removed and then occupied
in sequential steps. At each step, a set of $q$ unoccupied bonds is
randomly chosen. To each of these bonds a weight is assigned that is
proportional to the product of the size (number of sites) of the
clusters it would potentially join~\cite{Achlioptas09}.  In the case a
bond connects two sites in the same cluster, the weight is equal to
the square of the cluster size. From the random set of bonds chosen,
only the bond which has the smallest weight is occupied.  The
remaining bonds stay unoccupied, but can be selected again in later
steps. For $q=1$ we recover traditional percolation, while $q=2$
corresponds to the Achlioptas model~\cite{Achlioptas09}. We stop the
process when a spanning cluster is formed that connects the top and
bottom of the lattice. At this point we apply the burning
method~\cite{Herrmann84} to compute the mass of the spanning cluster
$M_{clus}$, the mass of the conducting backbone $M_{back}$, and the
number $M_{cut}$ of cutting bonds (bonds that would disconnect the
spanning cluster and stop conduction, if removed). The global
conductance $\sigma$ of the system is calculated by solving the set of
linear equations given by Kirchhoff's conservation law on each site.
Each of these quantities are averaged over at least 2000 realizations
for each value of $L$.

Typically for percolation two distinct criteria can be adopted to
define the critical occupation fraction $p_c$~\cite{Stauffer92}. Very
often the critical point is defined as the largest value of the bond
occupation probability $p$ where the largest cluster occupies a
vanishing fraction of the system in the thermodynamic limit, $L
\rightarrow \infty$. Alternatively, the critical point can be defined
as the smallest occupation fraction for which exists a cluster
connecting opposite sides of the lattice.  In the traditional
percolation model, $q=1$, these two definitions are
equivalent~\cite{Stauffer92}. For clarity, since we halt the
occupation of bonds when a spanning cluster is formed, we are here
examining the behavior of the system in terms of the later definition.
The scaling relations are used to identify the critical condition.
Moreover, the critical fraction and the exponents resulting from our
simulations for $q=2$ are consistent with the values reported in
Ref.~\cite{Radicchi09} for the Achlioptas model on square lattices.

\begin{figure}[t]
\begin{center}
\includegraphics*[width=8.0cm]{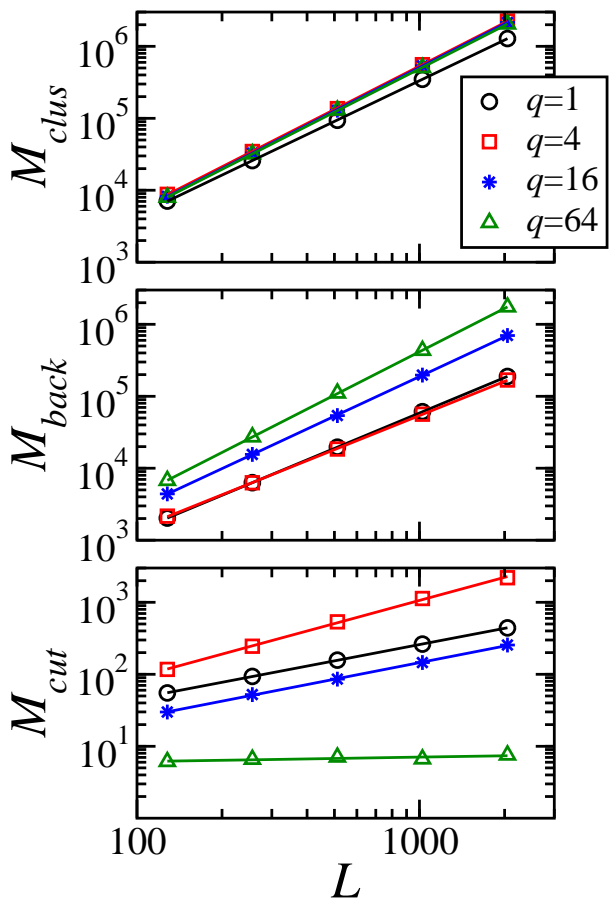}
\end{center}
\caption{(color online) Size dependence of the number of cutting
  bonds, $M_{cut}$, the number of sites in the conducting backbone,
  $M_{back}$, and the number of sites in the spanning cluster,
  $M_{clus}$. All quantities are averaged over at least 2000
  realizations precisely at the instant in which a percolation cluster
  is formed. For comparison, we present results for traditional bond
  percolation $q=1$, as well as for $q=4$, $16$ and $64$. The error
  bars are smaller than the symbols. In all cases, the measured
  quantities show a typical power-law dependence with the system size
  $L$ (solid lines). The corresponding values of the slopes of these
  curves are shown in Fig.~\ref{f.exp} as a function of $q$.}
\label{f.scal}
\end{figure}

\begin{figure}[t]
\begin{center}
\includegraphics*[width=8.0cm]{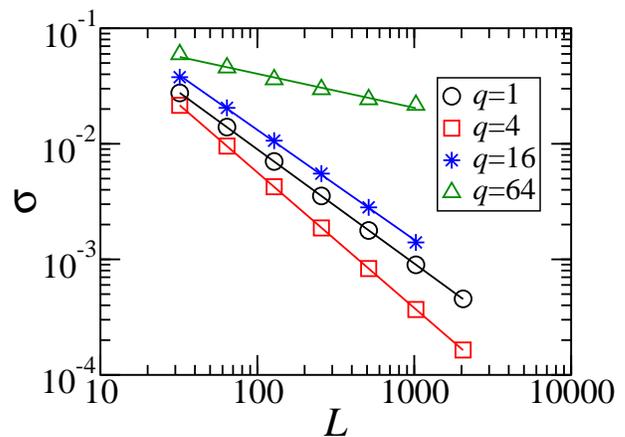}
\end{center}
\caption{(color online) Variation with size of the global conductance
  $\sigma$, for $q=1$, $4$, $16$ and $64$. As for the geometrical
  properties shown in Fig.~\ref{f.scal}, each point represents an
  average over at least 2000 realizations precisely at the critical
  percolation point. In all cases, the error bars are smaller than the
  symbols. The solid lines are the least-squares fits to the
  simulation data of power-laws, $\sigma\sim{L}^{-\mu}$. The
  corresponding exponents $\mu$ are shown in Fig.~\ref{f.exp} together
  with the results for other values of $q$.}
\label{f.scalcond}
\end{figure}

In Fig.~\ref{f.pic} we show typical clusters built with the
generalized EP model for $q=1$ (traditional percolation model), $q=2$
(Achlioptas model~\cite{Achlioptas09}), and $q=10$. As depicted in
Fig.~\ref{f.pic}a, the case $q=1$ results in a standard spanning
cluster with fractal dimension $d_{clus}=1.89$~\cite{Stauffer92},
which includes a conducting backbone of significantly smaller
dimension as a subset, $d_{back}=1.64$~\cite{Stauffer92}, and a few
cutting bonds, $d_{cut}=0.75$. In the Achlioptas model, $q=2$, the
spanning cluster incorporates a much larger fraction of the system,
while the conducting backbone becomes more tenuous with cutting bonds
appearing more frequently. As $q$ increases, both spanning cluster and
conducting backbone become larger and more compact, occupying most of
the lattice, with the number of cutting bonds being substantially
reduced. Surprisingly, the plots of the sample clusters shown in
Fig.~\ref{f.pic} suggest that some aspects of the system geometry
(e.g., the conducting backbone) do not necessarily change
monotonically as $q$ increases from one to very large values.
\begin{figure}[t]
\begin{center}
\includegraphics*[width=8.0cm]{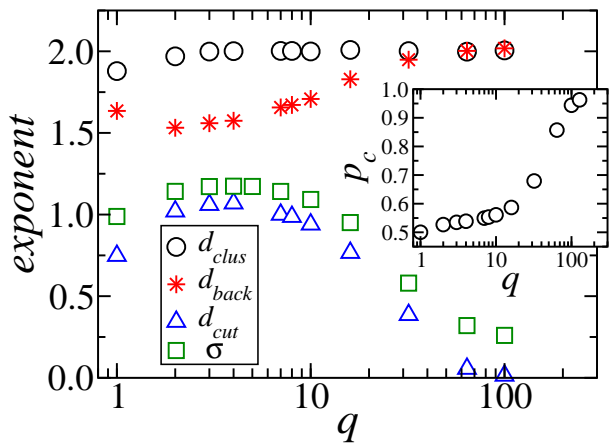}
\end{center}
\caption{(color online) The behavior of characteristic scaling
  exponents obtained for the quantities investigated in
  Figs.~\ref{f.scal} and \ref{f.scalcond}. As shown, the fractal
  dimension of the spanning cluster $d_{clus}$ grows with $q$,
  approaching $d_{clus}=2$, already at moderate values of $q$. In this
  way, the increase of the parameter $q$ leads to a percolation
  transition that is more abrupt.  Other exponents display an
  interesting non-monotonic dependence on $q$.  For instance, the
  fractal dimension of the conducting backbone $d_{back}$ is minimal
  at $q=2$, while the exponents for the conductance $\mu$ and number
  of cutting bonds $d_{cut}$ show a maximum around $q=4$. At large
  values of $q$, the conducting backbone becomes denser, $d_{back}=2$,
  with a conductance that is practically independent on the system
  size, $\mu \approx 0$. These results point to a first-order
  transition where percolation clusters are compact and occupy a large
  fraction of the lattice. Note also that the critical connectivity
  increases with $q$ approaching $p_c=1$ in the limit of large values
  of $q$, as shown in the inset.}
\label{f.exp}
\end{figure}

In order to characterize the critical properties of the model, we
perform finite-size scaling analysis. Figure~\ref{f.scal} shows the
average values of the quantities $M_{clus}$, $M_{back}$, and $M_{cut}$
at criticality against the system size $L$ for $q=1$, $4$, $16$ and
$64$. The scaling behavior with size for the conductivity $\sigma$ is
shown in Fig.~\ref{f.scalcond} for the same values of $q$. As
depicted, for any value of $q$, every measure depends on $L$ as a
power-law. In Fig.~\ref{f.exp} we show that the fractal dimension of
the spanning cluster increases monotonically from $d_{clus}=1.89$ at
$q=1$ to $d_{clus}=2.0$ at large $q$ values. The fractal dimension of
the conducting backbone goes from $d_{back}=1.64$ for $q=1$ to
$d_{back}=2.0$ at large values of $q$, but passing through a minimum
$d_{back}=1.52$ at $q=2$. On its turn, the fractal dimension
associated with the number (mass) of cutting bonds initially grows
from $d_{cut}=0.75$, reaches a maximum value of $d_{cut}=1.06$ at
$q=4$, and then decreases to values close to zero for sufficiently
large values of $q$. Similar to the exponent of the cutting bonds, the
exponent $\mu$ associated with the conductance also displays a maximum
around $q=4$. Starting from the standard percolation value of
$\mu=0.98$ at $q=1$, it reaches $\mu=1.18$ at $q=4$, and then
decreases monotonically, approaching $\mu=0$ in the limit of large $q$
values. For completeness, the inset of Fig.~\ref{f.exp} also shows the
behavior of the critical occupation fraction as it approaches the
limit $p_c=1$ as $q$ grows.

When the occupied bonds are selected from a large set, $q\to\infty$,
the behavior of the spanning clusters is equivalent to the one
observed in Ref.~\cite{Manna09}. Since the selection rule favors the
occupation of bonds connecting small clusters, the growth of the
spanning cluster is suppressed, leading to a significant increase in
the critical occupation fraction $p_c$, as compared to the standard
percolation value $p_c=0.5$. For very large values of $q$, the
spanning cluster is rather compact, saturated with internal bonds. 
All critical exponents obtained are consistent with homogeneous
networks, namely, $d_{clus}=d_{back}=2$, with the critical fraction
converging to its maximum possible value, namely, $p_c=1$. This
indicates that, for very large values of $q$, the transition is
similar to percolation in one dimension. Surprisingly, the behavior at
intermediate values of $q$ is the least favorable for conduction, that
is, the conductance decreases faster with system size. At this
condition, one also observes that the fractal dimension of the
backbone $d_{back}$ is minimal and the exponent controlling the
scaling of the cutting bonds is maximal. As we discuss next, the
reason for this behavior is closely related to the product rule
selection criteria as well as the degree of heterogeneity of the
cluster size distribution prior to the critical point.

During the EP process, bonds selected from the $q$-set can be either
merging bonds, i.e., bonds that connect sites belonging to two
distinct clusters, or internal bonds, i.e., bonds connecting two sites
inside the same cluster. Note that only the inclusion of internal
bonds can form loops in the system. For any value of $q>1$, however,
the inclusion of internal bonds in the larger clusters is hindered.
Larger clusters have more internal bonds than smaller clusters, so it
is more likely that the internal bonds included in the selection set
already belong to the largest clusters in the system. However, since
these bonds have the largest possible weight, their occupation is less
probable. Now we argue that the fractal dimension of the backbone is
dominated by the proportion of loops. For very large values of $q$,
all clusters have more or less the same size and, at $p_{c}$, two
nearly compact clusters are joined to form the spanning cluster, which
means that a considerable fraction of the resulting backbone consists
of loops. Therefore, the fractal dimension of the backbone for large
$q$ is close to that of the spanning cluster itself, both approaching
the topological dimension of the lattice in the limit $q\to\infty$.
For intermediate values of $q$, however, the cluster size distribution
is not that strongly monodisperse and, due to the selection rule, the
connection of clusters of intermediate sizes dominates over the
closing of loops in the larger clusters. At $p_{c}$ the spanning
cluster is typically created by joining two of the larger clusters.
Thus the backbone will have a particularly low proportion of loops and
therefore its fractal dimension will be low.  This explains the
observed minimum.

In summary, we have studied the transport properties of percolation
networks built from a generalization of the Achlioptas
model~\cite{Achlioptas09}, where a bond is occupied from a set of
$q\ge{1}$ randomly selected bonds, according to the product rule of
cluster sizes that they would potentially connect. Our results
indicate that, for any value of $q$, the mass of the conducting
backbone, the total mass of cutting bonds, and the global conductance
of networks at the critical point, all scale as power-laws with system
size. Moreover, the scaling exponents converge to the exponents of a
fully occupied square lattice, in the limit of very large values of
$q$. This behavior is consistent with a first-order phase transition.
Interestingly, we observe that systems with intermediate values of $q$
display the worst conductive performance.  We argue that conduction
becomes difficult for these values of $q$ because the formation of
loops in the spanning cluster is prevented, resulting in smaller
conducting backbones.

We thank the Brazilian Agencies CNPq, CAPES, FUNCAP and FINEP, the
FUNCAP/CNPq Pronex grant, and the National Institute of Science and
Technology for Complex Systems in Brazil for financial support.


\begin{thebibliography}{1} 

\bibitem{Chalupa79} J. Chalupa, P. L. Leath, and G. R. Reich, J.
  Phys. C \textbf{12}, L31 (1979); R. Dickman and T. Tome, Phys. Rev.
  A \textbf{44}, 4833 (1991); C. Moukarzel, P. M. Duxbury, and P. L.
  Leath, Phys. Rev. Lett. \textbf{78}, 1480 (1997); M. A. Knackstedt,
  M. Sahimi, and A. P. Sheppard, Phys. Rev. E \textbf{61}, 4920
  (2000); R. Parshani, S. V. Buldyrev, and S. Havlin, Phys. Rev.
  Lett. \textbf{105}, 048701 (2010).

\bibitem{Achlioptas09} D. Achlioptas, R. M. D'Souza, and J. Spencer,
  Science \textbf{323}, 1453 (2009).

\bibitem{Ziff09} R. M. Ziff, Phys. Rev. Lett. \textbf{103}, 045701
  (2009); R. M. Ziff, arXiv:0912.1060v3.

\bibitem{Radicchi09} F. Radicchi and S. Fortunato, Phys. Rev. Lett.
  \textbf{103}, 168701 (2009); Phys. Rev. E \textbf{81}, 036110
  (2010).

\bibitem{Friedman09} E. J. Friedman and A. S. Landsberg, Phys. Rev.
  Lett. \textbf{103}, 255701 (2009).

\bibitem{Costa10} R. A. da Costa, S. N. Dorogovtsev, A. Goltsev, and
  J. F. F. Mendes, arXiv:1009.2534v2.

\bibitem{Rozenfeld10} H. D. Rozenfeld, L, K. Gallos, and H. A. Makse,
  Eur. Phys. J. B \textbf{75}, 305 (2010).

\bibitem{DSouza10} R. M. D'Souza and M. Mitzenmacher, Phys. Rev.
  Lett. \textbf{104}, 195702 (2010).

\bibitem{Cho10} Y. S. Cho, J. S. Kim, J. Park, B. Kahng, and D. Kim,
  Phys. Rev. Lett. \textbf{103}, 135702 (2009); Y. S. Cho, B. Kahng,
  and D. Kim, Phys. Rev. E \textbf{81}, 030103 (2010).

\bibitem{Moreira10} A. A. Moreira, E. A. Oliveira, S. D. S. Reis, H.
  J. Herrmann, and J. S. Andrade, Phys. Rev. E \textbf{81}, 040101
  (2010).

\bibitem{Araujo10} N. A. M. Araujo and H. J. Herrmann, Phys. Rev. Lett.
  \textbf{105}, 035701 (2010).

\bibitem{Stauffer92} D. Stauffer and A. Aharony, {\it Introduction to
    Percolation Theory}, (Taylor \& Francis, London, 1992).

\bibitem{Sahimi94} M. Sahimi, \textit{Applications of Percolation
    Theory} (Taylor \& Francis, London, 1994).

\bibitem{Last71}B. J. Last and D. J. Thouless, Phys. Rev. Lett.
  \textbf{27}, 1719 (1971); B. I.  Halperin, S. Feng, and P. N. Sen,
  Phys.  Rev. Lett. \textbf{54}, 2391 (1985); Y. Meir, Phys. Rev.
  Lett.  \textbf{83}, 3506 (1999); C.  Grimaldi and I. Balberg, Phys.
  Rev. Lett.  \textbf{96}, 066602 (2006).

\bibitem{Stanley84} H. E. Stanley and A. Coniglio, Phys. Rev. B
  \textbf{29}, 522 (1984); J. S. Andrade, D. A.  Street, T. Shinohara,
  Y. Shibusa, and Y. Arai, Phys. Rev. E \textbf{51}, 5725 (1995); J.
  S. Andrade, M. P.  Almeida, J. Mendes Filho, S. Havlin, B. Suki, and
  H. E. Stanley, Phys. Rev. Lett. \textbf{79}, 3901 (1997); M. Sahimi,
  M. Hashemi, and J. Ghassemzadeh, Physica A \textbf{260}, 231
  (1998); A. Klemm, R. Kimmich, and M. Weber, Phys. Rev. E \textbf{63}, 
  041514 (2001).

\bibitem{Manna09} S. S. Manna and A. Chatterjee, arXiv:0911.4674v1.

\bibitem{Herrmann84} H. J. Herrmann, J. Phys. A {\bf 17}, L261 (1984).

\end{thebibliography}
\end{document}